\documentclass{article}
\usepackage{ijcai24}

\usepackage{times}
\usepackage{soul}
\usepackage{url}
\usepackage[hidelinks]{hyperref}
\usepackage[utf8]{inputenc}
\usepackage[small]{caption}
\usepackage{graphicx}
\usepackage{amsmath}
\usepackage{amsthm}
\usepackage{booktabs}
\usepackage{algorithm}
\usepackage{algorithmic}
\usepackage[switch]{lineno}
\usepackage{multirow}
\usepackage{dsfont}
\usepackage{amssymb}
\usepackage{xcolor}
\usepackage{tabularx}
\usepackage{float}

\title{Improving Paratope and Epitope Prediction by Multi-Modal Contrastive Learning and Interaction Informativeness Estimation}

\author{
Zhiwei Wang$\textsuperscript{\rm 1}^{\dagger}$\and
Yongkang Wang$\textsuperscript{\rm 1}^{\dagger}$\and
Wen Zhang$\textsuperscript{\rm 1,\rm 2 \rm 3}^{\ast}$\\
\affiliations
\textsuperscript{\rm 1}College of Informatics, Huazhong Agricultural University, Wuhan 430070, China\\
\textsuperscript{\rm 2}Hubei Key Laboratory of Agricultural Bioinformatics, Huazhong Agricultural University, Wuhan 430070, China\\
\textsuperscript{\rm 3}Engineering Research Center of Intelligent Technology for Agriculture, Ministry of Education, Wuhan 430070, China\\
\emails
\{wangzhiwei, wyky481\}@webmail.hzau.edu.cn,
zhangwen@mail.hzau.edu.cn
}

\begin{document}

\maketitle

\begin{abstract}
Accurately predicting antibody-antigen binding residues, i.e., paratopes and epitopes, is crucial in antibody design. However, existing methods solely focus on uni-modal data (either sequence or structure), disregarding the complementary information present in multi-modal data, and most methods predict paratopes and epitopes separately, overlooking their specific spatial interactions. In this paper, we propose a novel \textbf{M}ulti-modal contrastive learning and \textbf{I}nteraction informativeness estimation-based method for \textbf{P}aratope and \textbf{E}pitope prediction, named \textbf{MIPE}, by using both sequence and structure data of antibodies and antigens. MIPE implements a multi-modal contrastive learning strategy, which maximizes representations of binding and non-binding residues within each modality and meanwhile aligns uni-modal representations towards effective modal representations. To exploit the spatial interaction information, MIPE also incorporates an interaction informativeness estimation that computes the estimated interaction matrices between antibodies and antigens, thereby approximating them to the actual ones. Extensive experiments demonstrate the superiority of our method compared to baselines. Additionally, the ablation studies and visualizations demonstrate the superiority of MIPE owing to the better representations acquired through multi-modal contrastive learning and the interaction patterns comprehended by the interaction informativeness estimation.
\end{abstract}

\section{Introduction}
\begin{figure}[t]
    \center{\includegraphics[width=8cm]{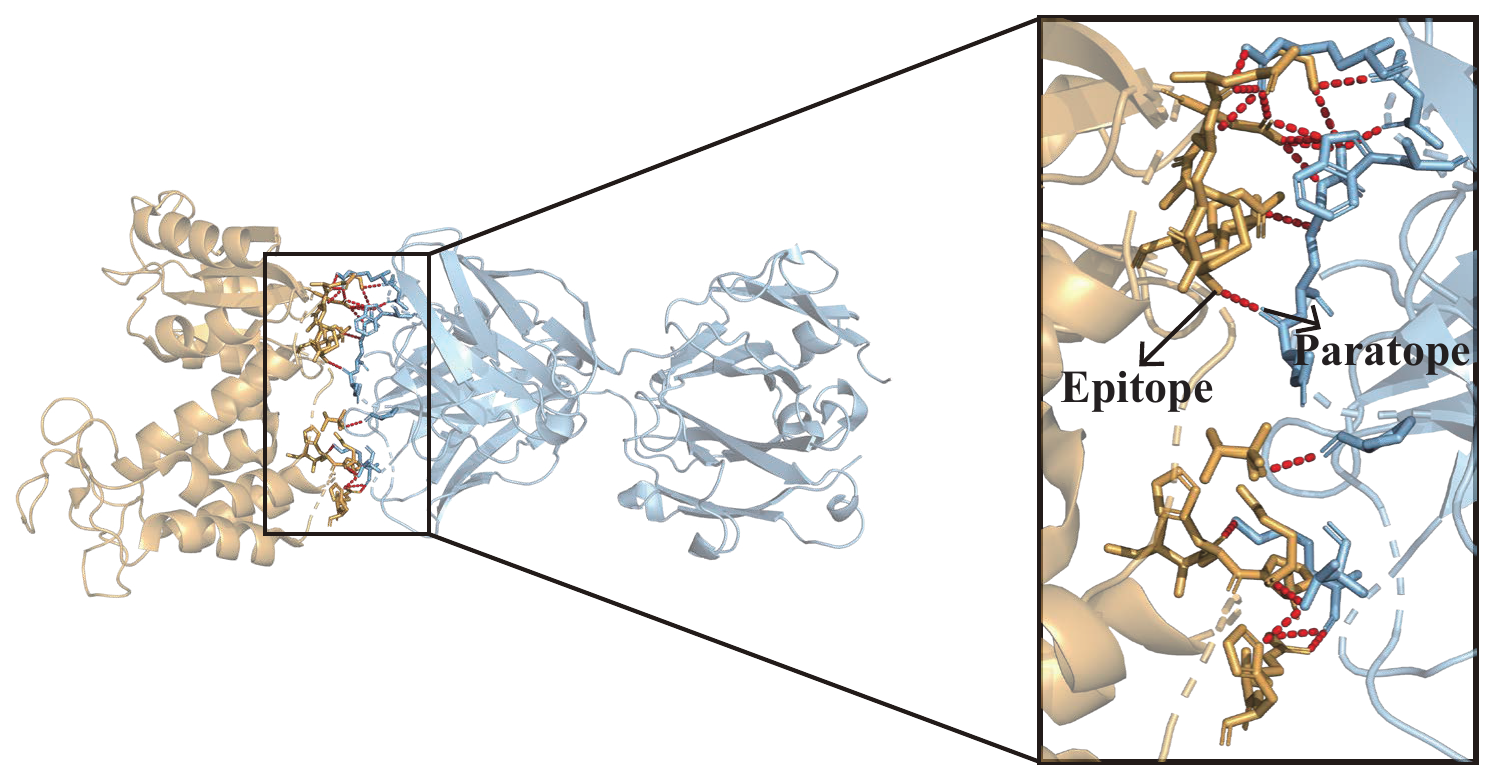}} 
    \caption{The interaction map between HexaBody-CD38 Fab and CD38 (PDB: 8BYU).}
\end{figure}

An antigen (Ag) is a typical protein that elicits an immune response in the human body, while antibodies (Ab) are flexible Y-shaped proteins \cite{Antibodies2020}. The antigen interacts with the antibody and then initiates immune responses. As shown in Figure 1, these binding residues on antibodies and antigens are known as paratopes and epitopes \cite{BcellPNAS}, respectively. The identifying epitopes help to construct epitope-based antibodies while confirming paratopes allows precise mutation of binding regions, optimizing properties like affinity \cite{SA2022}. Traditionally, researchers employed wet experiments such as X-ray crystallography, nuclear magnetic resonance spectroscopy, and cryoelectron microscopy for identifying paratopes and epitopes \cite{Xray,NMR,CryoElectron}. However, these approaches are often costly, time-consuming, and struggle to cope with the scale of rapidly emerging high-throughput data \cite{hits}.

Due to the special structure of antibodies, CDR regions are more prone to conformational changes, resulting in more complex interaction patterns, and the general methods for proteins cannot perform well on paratope and epitope prediction. In recent years, plenty of computational methods have been developed for paratope and epitope prediction. These methods are roughly classified into three categories: machine learning-based methods, deep learning-based methods, and pre-trained language model-based methods. The machine learning-based methods \cite{proABC,3DZernike} extracted the hand-crafted features of antibodies and antigens, and fed the features into classifiers (e.g., SVM, random forest) to predict paratopes and epitopes. The deep learning-based methods \cite{AGFastParapred,NetBCE} utilize multi-layer neural networks, including convolutional neural networks (CNNs), and equivariant graph neural networks (EGNNs), to extract representations from input data. Currently, pre-trained language models (PLMs) like ESM-1b, and ProtTrans are trained on extensive protein sequences across various downstream tasks, including the paratope and epitope prediction \cite{BepiPred-3.0}.

The existing methods have made great progress in paratope and epitope prediction. However, given the specificity of antibodies, the prediction of paratopes and epitopes still faces many challenges. In particular, the scarcity of labeled paratopes and epitopes is always a problem to be solved, posing a challenge in extracting abundant information from limited data. Additionally, paratope and epitope prediction contains three tasks: single paratope/epitope prediction, and joint paratope-epitope prediction. However, existing methods usually focus on one of them and do not consider all three tasks.

Multi-modal information from sequences and structures could provide richer complementary knowledge on limited data. Effectively leveraging the multi-modal information of antibodies and antigens may improve paratope and epitope prediction, but still remains challenging. Antibodies (antigens) have multiple paratopes (epitopes), leading to potential binding with different counterparts. However, existing methods extract representations solely from antibodies (antigens) to predict paratopes (epitopes), but they fail to exploit the information on how epitopes are specifically bound to paratopes, especially in single paratope/epitope prediction. Therefore, incorporating the antibody-antigen interaction information into models could further enhance the performances of paratope and epitope prediction.

\begin{figure*}[h]
    \center{\includegraphics[width=17.8cm]{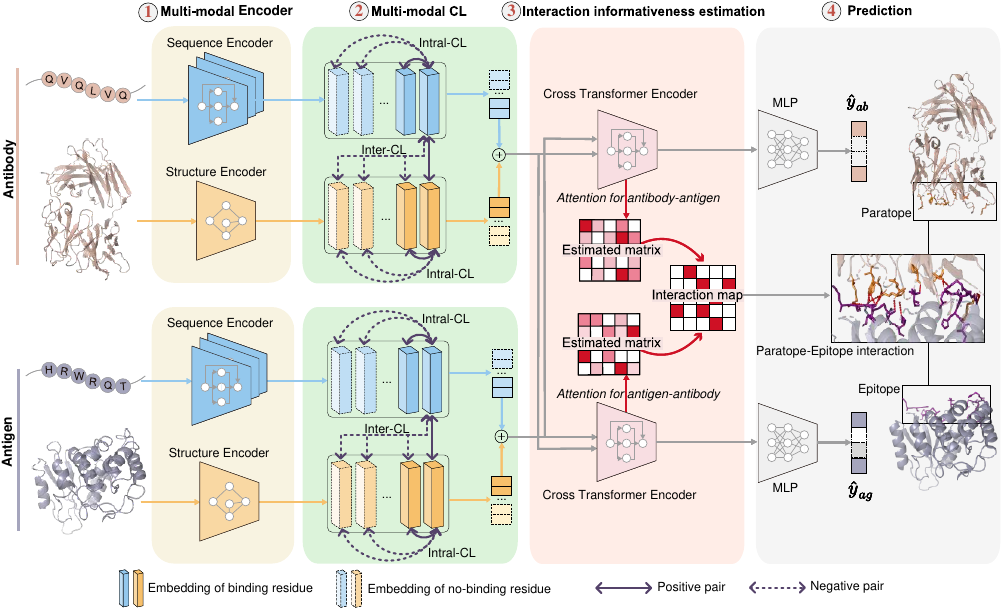}} 
    \caption{The overview of our proposed MIPE. The yellow part is the multi-modal encoders, the green part is the multi-modal contrastive learning, the red part is the interaction informativeness estimation, and the gray part is the prediction.}
\end{figure*} 

To address these challenges, we propose a novel \textbf{M}ulti-modal contrastive learning and \textbf{I}nteraction informativeness estimation-based method for \textbf{P}aratope and \textbf{E}pitope prediction, named \textbf{MIPE}. Specifically, we introduce multi-modal contrastive learning (CL) to effectively integrate the modal information from sequences and structures. Within each modality, we deploy the intra-CL to maximize representations of binding and non-binding residues; between different modalities, we deploy the inter-CL to align uni-modal representations towards effective modal representations, reducing the impact of noise. Furthermore, we devise a knowledge-driven module named interaction informativeness estimation, which utilizes multi-head attention layers to compute the attention matrices for antibodies and antigens, aiming to capture the specific interaction patterns between them. 

In summary, the main contributions of this paper are described as follows:
\begin{itemize}
    \item We introduce a multi-modal contrastive learning strategy that learns the representation of paratopes and epitopes by applying intra-CL and inter-CL to multi-modality sequence and structure data. 
   \item We devise an interaction informativeness estimation to approximate the estimated matrix to the actual antibody-antigen interaction map, which guides the model to learn the specific patterns of the antibody-antigen interactions.
    \item Extensive experiments show that MIPE achieves competitive performance in paratope and epitope prediction, and also results in promising performance even utilizing AlphaFold2-generated structures as substitutes for real structures which are unavailable.
\end{itemize}

\section{Related Work}
For the paratope and epitope prediction, existing computational methods focus on three tasks: single paratope prediction, single epitope prediction, and joint paratope-epitope prediction.
For instance, Parapred \cite{Parapred} employs the CNN and BiLSTM networks and integrates features from both the local residue domain and the entire antibody sequence for paratope prediction. Paragraph \cite{Paragraph} utilizes EGNNs to extract 3D representations of the antibody's CDR regions for paratope prediction. 
For single epitope prediction, NetBCE \cite{NetBCE} extracts sequence features of antigens and employs BiLSTMs and CNNs to learn the representations to predict epitopes. BepiPred-3.0 \cite{BepiPred-3.0} leverages ESM-2 to obtain sequence representation of antigens and subsequently employs various classifiers (e.g., FFNN and CNN) for epitope prediction.
For joint paratope-epitope prediction, \cite{ICML2020} proposed a cross-attention network to encode both the antigen and antibody sequences. Epi-EPMP \cite{ICML2021} designed a graph attention neural network, incorporating the encoding of the structural graphs of both the antibody and antigen. In general, an ideal prediction method is deemed to perform well on all three tasks.

Although existing methods produce good performances, they typically concentrate on uni-modal data (sequences or structures) while overlooking the information across multi-modal data. Studies have shown that multi-modal data can offer complementary information \cite{MMcolearning}, and such multi-modal based approaches have been successfully applied in various fields, e.g., drug-target interaction and protein-protein interaction \cite{multimodalppi,multimodaldti,MMCD}. To fill the gap, we propose a multi-modal contrastive learning strategy for the sequence-structure modality fusion to learn better representation for the paratope and epitope prediction. 

As mentioned above, the antibody-antigen interactions reflect how paratopes and epitopes are bound, which could benefit the paratope and epitope prediction, but such explicit interaction information hidden in the known antigen-antibody complexes has not been exploited in the previous studies \cite{ICML2020,ICML2021}. Moreover, the lack of explicit binding information from structures may undermine the performance of the prediction model and also hinder the model from proficiently analyzing the interaction patterns. To address this issue, we present an interaction informativeness estimation strategy to facilitate the model's learning of interaction patterns, guided by the actual antibody-antigen interaction map.

\section{Methodology}
In this section, we first formulate the problem of paratope and epitope prediction. Subsequently, we introduce the components of the proposed method MIPE, which includes multi-modal encoders, multi-modal contrastive learning, and interaction informativeness estimation, as illustrated in Figure 2.

\subsection{Problem Formulation}
An antibody or antigen in the antibody-antigen complex can be represented as a sequence-structure tuple $(\textit{S},\textit{C})$. $\textit{S}=[s_i]_{i=1}^{M}$ stands for the residue arrangements where $s_i$ is the type of residue $i$, and $\textit{C}=[c_i]_{i=1}^{M}$ stands for the backbone coordinates where $c_i$ is a Cartesian coordinate of the C$_\alpha$ atom in residue $i$. Binding residues are determined by identifying residues within a Euclidean distance threshold between the antibody and the antigen, and paratopes/epitopes refer to binding residues on the antibody/antigen.

Given a set of known antibody-antigen complexes, we can obtain the dataset containing sequence and structure data of antibodies and antigens, the annotated paratopes and epitopes, and the interaction map. The objective of our study is to train a model using the dataset and then apply it to predict paratopes and epitopes of new antibodies and antigens.

\subsection{Multi-Modal Encoders}
To build our model, the first step is to characterize the sequences and structures of antibodies and antigens, by using a sequence encoder and a structure encoder.

For sequences, we initialize them as residue-type features (one-hot encoding with 20 dimensions): $x^s \in \mathbb{R}^{M \times 20}$. Then, we implement the sequence-based encoder to obtain the embedding from $x^s$, formalized as follows:
\begin{equation}
      h^s=\mathbb{F}_s(x^s)
\end{equation}
where $\mathbb{F}_s$ consists of a pre-trained network\cite{ESM1b,AbLang}, a dilated convolutional neural network\cite{DCNN}, and a recurrent network\cite{BiLSTM} consecutively.

For structures, we build the structure encoder as follows. First, we construct a 3D graph based on residue coordinates and add three different types of edges to the graph. If the sequential distance between the $i$-th residue and the $j$-th residue is below a predefined threshold $d_{seq}$, the edge between these two residues is considered a sequential edge. If the Euclidean distance between two residues is smaller than a threshold $d_{radius}$, the edge between them is regarded as the radius edge. To ensure the density of spatial edges is comparable among different graphs, for each residue, $s_i$, the $k$ residues with the smallest Euclidean distances to it are considered its $K$-nearest neighbors, and the edges between them are considered as $K$-nearest neighbor edges. Herein, we set $d_{seq}=3$, $d_{radius}=10$Å, $K=8$. For each node in the graph, we construct an initial 62-dimensional physicochemical feature: $x^c\in\mathbb{R}^{M\times 62}$ (see details in Appendix 1). Following \cite{GearNet}, we employ the equivariant graph convolutional layers to learn and update the graph:
\begin{equation}
h^{l+1},s^{l+1}=\mathrm{EGCL}[h^l,s^l,\mathcal{E}]
\end{equation}
where $l$ denotes the number of layers. $h=\{h^c_1,...,h^c_M\}$ represents the set of node embeddings, which are initialized with $x^c$. $s=\{s_1,...,s_M\}$ is the set of residue coordinates. $\mathcal{E}=(e_{ij})$, $e_{ij}$ denotes the feature of the edge between node $i$ and $j$. Finally, we apply an MLP over the updated embeddings of all nodes to produce the output embeddings for the residues of antibodies and antigens.

\subsection{Multi-Modal Contrastive Learning}
After encoding sequences and structures, it becomes crucial to fuse their embeddings for the paratope and epitope prediction. Mutual information (MI) offers an effective measure for assessing nonlinear dependencies between variables, which compels the alignment of modal data and facilitates the sharing of crucial information \cite{MI}. Along this line, we bring in contrastive learning (CL) to align sequences and structures by maximizing their MI in the embedding space \cite{CL1}. Specifically, we devise the multi-modal CL strategies with intra-CL and inter-CL, as follows:

\paragraph{Intra-CL.} To better discriminate residues between binding and non-binding in both sequences and structures, it is desirable for the binding residues to cluster together while being distant from non-binding ones in the embedding space. To do so, in a given antibody/antigen with length $M$, we define the binding residue $i$ as the anchor, another binding residue $j$ as the positive instance, and the non-binding residue $k$ is regarded as a negative instance. Then, we maximize the MI between binding residues while minimizing the MI between the binding and non-binding residues at the sequence and structure levels. The objective of intra-CL is to minimize the following InfoNCE-based \cite{chen2020simple} loss function:
\begin{equation}
\begin{aligned}
    \mathcal{L}_{intra} & =-\frac{1}{M}\sum_{i=1,j\neq{i}}^{M}\left({\rm log}\frac{E(h_{i}^s,h_{j}^s)}{\sum_{k=1}^{M}\mathds{1}_{[k\neq{i}]}E(h_{i}^s,h_{k}^s)} \right.
\\&\left. + {\rm log}\frac{ E(h_{i}^c,h_{j}^c) }{\sum_{k=1}^{M}\mathds{1}_{[k\neq{i}]}E(h_{i}^c,h_{k}^c)}\right)
\end{aligned}
\end{equation}
where $h_{i}^s$/$h_{i}^c$ represent the sequence/structure representation of residue $i$ (refers to the anchor), $j$ means the residue indices consistent with the label of residue $i$; $\mathds{1}_{[k\neq{i}]}\in\left\{0,1\right\}$ is an indicator function evaluating to 1 if $k\neq{i}$; $E(\cdot,\cdot)$ is the cosine similarity function with the temperature coefficient to measure the MI score between two variables.

\paragraph{Inter-CL.} To acquire dependable multi-modal representations of antibodies/antigens, it is preferable to reduce the prediction bias in different modalities and strengthen their consistency. Inspired by \cite{zou2023unis}, we bring in an inter-CL to encourage modalities with both correct uni-modal predictions to share a stronger correspondence. Specifically, for the learned sequence and structure representations of residues, we utilize two pre-trained probabilistic discriminators for calculating their pseudo labels:
\begin{equation}
    p_{i}^{s}=f^{s}(h_{i}^{s}),\quad p_{i}^{c}=f^{c}(h_{i}^{c}).
\end{equation}
where $f^{s}$/$f^{c}$ is the sequence-based/structure-based discriminator (see details in Appendix 1), $p_{i}^{s}$/$p_{i}^{c}$ stands for the obtained pseudo labels of residue $i$. 

If the obtained labels of both sequence and structure ($p_{i}^{s}$, $p_{i}^{c}$) are consistent with the original labels, corresponding $h_{i}^{s}$ and $h_{i}^{c}$ are regarded as positive pairs, forming the set of all positive pairs (denoted as $\mathds{P}$). If the obtained labels of both sequence and structure are inconsistent with the original labels, they are defined as negative pairs, denoted as $\mathds{N}$. Additionally, when only one of the predicted binding labels is consistent with its original label, these two modal representations are defined as a semi-positive pair $\mathds{S}$. The objective is to minimize the following loss function:
\begin{equation}
\mathcal{L}_{inter}=-{\rm log}\frac{\textstyle \sum_{i\in \mathds{P},\mathds{S}} E(h_{i}^s,h_{i}^c)}{\textstyle \sum_{j\in \mathds{B}} E(h_{j}^s,h_{j}^c)) + \textstyle \sum_{i,j \in \mathds{B} \wedge i \neq j} E(h_{i}^s,h_{j}^c))}
\end{equation}
where $\mathds{B} = \mathds{P} \cup \mathds{S} \cup \mathds{N}$. In addition to $\mathds{N}$, different modal embeddings of distinct residues need to separate their representations; therefore, when $i \neq j$, they are also considered negative pairs.

Above inter-CL can optimize the similarity of positive and semi-positive pairs toward a higher value while optimizing the similarity of negative pairs toward a smaller value. The reason behind this design is twofold: (1) for samples with incorrect predictions in both modalities, we aim to enhance the dissimilarity between their uni-modal representations to obtain more complementary representations; (2) for samples with mutually exclusive predictions, we encourage the alignment of ineffective modality with the effective one by leveraging uni-modal predictions as supervision signals.

Furthermore, we fuse the sequence and structure representations through point-wise addition: 
\begin{align}
    h=h^{s} \oplus h^{c}
\end{align}

\subsection{Interaction Informativeness Estimation}
How epitopes are specifically bound to paratopes is a crucial question overlooked by existing research. Therefore, incorporating the interaction between antibodies and antigens could further enhance the performances of paratope and epitope prediction. Here, we design an interaction informativeness estimation (IIE) to incorporate the interaction maps into the model prediction.

First, we deploy the transformer encoder to estimate interaction maps between antibodies and antigens from the learned representations \cite{attention}. The estimated matrix $A$ can be defined by the following cross-attention layer: 
\begin{equation}
A=\operatorname{softmax}\left(\frac{Q K^T}{\sqrt{d}}\right)V
\end{equation}
where the antibody (or antigen) representations of sequence and structure are integrated as the query $Q=h_{ab}$ (or $Q=h_{ag}$), the antigen (or antibody) representations serve as the key and value $K=V=h_{ag}$ (or $K=V=h_{ab}$), $d$ denotes the representation dimension of residues. After that, we can obtain the antibody-oriented and antigen-oriented attention matrices: ${A}_{ab} \in \mathbb{R}^{M\times N}$ and ${A}_{ag} \in\mathbb{R}^{N\times M}$, where ${A}_{ab}(i,j)$/${A}_{ag}(i,j)$ represents the binding score between the $i$-th residue on the antibody/antigen and the $j$-th residue on the antigen/antibody.

Both $A_{ab}$ and $A_{ag}$ encapsulate the estimated antibody-antigen interaction matrix. Our objective is to approximate these two estimated matrices towards the actual antibody-antigen interaction map, facilitating the model in capturing specific interaction patterns and thereby enhancing overall performance. This process enables the model to learn and represent the complexities inherent in antibody-antigen interactions more effectively. With the actual antibody-antigen interaction map $\hat{A} \in\mathbb{R}^{M\times N}$, we minimize their distance by the following BCE loss:
\begin{equation}
\begin{aligned}
\mathcal{L}_{IIE}=-&\left(\hat{A} \log A_{a b}+(1-\hat{A}) \log \left(1-A_{a b}\right) \right. + \\
              &\left. \hat{A} \log A_{a g}^T+(1-\hat{A}) \log \left(1-A_{a g}^T\right)\right)
\end{aligned}
\end{equation}

\subsection{Prediction and Model Training}
For paratope and epitope prediction, the embedding of each residue on the antibody and antigen ($h_{ab}$ and $h_{ag}$) derived from the transformer encoders are individually inputted into MLP-based predictors $\mathcal{F}_{ab}$ and $\mathcal{F}_{ag}$. These predictors assess the probability of a residue being the paratope/epitope: 
\begin{align}
     \hat{y}_{ab}=\mathcal{F}_{ab}(h_{ab}),\quad \hat{y}_{ag}=\mathcal{F}_{ag}(h_{ag})
\end{align}
The supervised loss of the paratope prediction can be formulated as follows:
\begin{equation}
     \mathcal{L}_{ab} =  -\frac{1}{M}\sum\limits_{i=1}^{M} \left( y_{ab_i}\text{log}\left(\hat y_{ab_i} \right)+\left(1- y_{ab_i}\right)\text{log}\left(1-\hat y_{ab_i}\right) \right)
\end{equation}
where $y_{ab_i}$ indicates the ground-truth label of the $i$-th residue of the antibody. Similarly, the loss for the epitope prediction is defined as $\mathcal{L}_{ag}$. 

By combining the loss functions for Intra-CL, Inter-CL, IIE, the paratope prediction, and the epitope prediction, the overall objective function for model training is defined as:

\begin{align}
     \mathcal{L} = & \alpha\mathcal{L}_{CL} + \beta\mathcal{L}_{IIE} + \gamma (\mathcal{L}_{ab} + \mathcal{L}_{ag}).
\end{align}
where $\alpha,\beta,\gamma$ represents a hyperparameter to balance the contributions of different tasks.

\section{Experiments}
\subsection{Experimental Settings}
\paragraph{Datasets.} From the publicly accessible Structural Antibody Database (SAbDab), we collected a total of 7571 antibody-antigen complexes, with the sequence data in FASTA format and structural data in PDB format. Following previous studies \cite{PECAN}, we used CD-HIT \cite{cdhit} to remove high-homology antibody and antigen sequences with the thresholds of $95\%$ and $90\%$ sequence identity, respectively. Subsequently, we excluded antibodies and antigens with any residue type rather than 20 naturally occurring types. Finally, we compiled a dataset consisting of 626 binding antibody-antigen pairs, including their sequences, structures, and corresponding interaction maps. Noteworthy, antibodies primarily bind to antigens through their CDR regions. Most researchers use Euclidean distance to define paratopes and epitopes, and we follow the usual way in our dataset: within the CDR regions/antigen, a residue is labeled as a paratope/epitope if the Euclidean distance between its backbone atom and any backbone atom on the other antigen/CDR regions is less than 4.5 Å.

\paragraph{Baselines.} We compared our MIPE with the following methods for the paratope and epitope prediction at three different tasks. For the single paratope prediction, the Parapred \cite{Parapred}, AG-Fast-Parapred \cite{AGFastParapred}, PECAN \cite{PECAN} and Paragraph \cite{Paragraph} are listed as the baselines. For the single epitope prediction, we took BepiPred-3.0 \cite{BepiPred-3.0}, NetBCE \cite{NetBCE} and PECAN \cite{PECAN} as the baselines. PesTo \cite{PesTo} is a method used to predict protein binding sites and is applicable to both the single prediction task for paratopes and epitopes. Additionally, for the joint paratope-epitope prediction, we leveraged the \cite{ICML2020} and Epi-EPMP \cite{ICML2021} as the compared methods.

\paragraph{Evaluation metrics.} We randomly split the dataset, allocating $90\%$ binding antibody-antigen pairs for cross-validation (CV) and reserving the remaining $10\%$ for an independent test set. For the CV set, we implemented the 5-fold CV. Furthermore, we performed independent testing, in which the model was trained on the CV set and evaluated on the separate and previously untouched independent test set. Following previous studies, we employed three evaluation metrics to evaluate the performance of paratope and epitope prediction. Due to the class imbalance in binding and non-binding residues, the area under the precision-recall curve (\textbf{AUPR}) is the primary metric for measuring the predictive performance. We also reported the metrics, such as the area under the receiver operating characteristic curve (\textbf{AUC}) and the Matthews correlation coefficient (\textbf{MCC}), consistent with prior research, to facilitate the comparisons. Notably, the reported values are aggregated across five random seeds. Details about the datasets, baselines, metrics, and implementations can be found in Appendix 2. Our code, data, and appendix are available on GitHub ({\url{https://github.com/WangZhiwei9/MIPE}})

\subsection{Performance Comparison}
\begin{table*}[ht]
    \centering
    \renewcommand\arraystretch{1.2}
    \small
    \setlength{\tabcolsep}{2.1mm}{
    \begin{tabular*}{1\linewidth}{clcccccc}
        \hline
        \multirow{2}{*}{Task}  & \multirow{2}{*}{Method}  & \multicolumn{3}{c}{Paratope}    &  \multicolumn{3}{c}{Epitope}  \\
         ~  &  ~  &  AUC  &  AUPR  &  MCC  &  AUC  &  AUPR  &  MCC \\
        \hline
         ~  &  Parapred     & 0.868±0.002     & 0.652±0.002     & 0.503±0.001     & -     & -     & - \\
        ~  &  AG-Fast-Parapred     & 0.883±0.004     & 0.612±0.003     & 0.548±0.003     & -     & -     & - \\
        Single  &  PECAN     & \underline{0.915±0.000}     & 0.713±0.001     & \textbf{0.558±0.001}     & -     & -     & - \\
        Paratope  &  Paragraph     &\textbf{0.927±0.000}     & 0.650±0.000     & 0.488±0.001     & -     & -     & -   \\
        Prediction  &  PesTo     & 0.856±0.001     & \underline{0.721±0.001}     & 0.433±0.003     & -     & -     & -   \\
        ~  &  MIPE     & \textbf{0.927±0.000}    & \textbf{0.741±0.000}     & \underline{0.554±0.001}     & -     & -     & -  \\
        ~  &  MIPE (AlphaFold2)     & {0.910±0.000}    & {0.723±0.000}     & {0.531±0.001}     & -     & -     & -  \\
        \hline
        ~  &  BepiPred-3.0     & -     & -     & -     & 0.774±0.001     & 0.395±0.004     & \textbf{0.292±0.002} \\
        ~  &  NetBCE     & -     & -     & -     & \underline{0.845±0.000}     & \textbf{0.517±0.000}     & 0.239±0.002 \\
        Single  &  PECAN     & -     & -     & -     & 0.637±0.001     & 0.201±0.004     & 0.150±0.003 \\
        Epitope  &  PesTo     & -     & -     & -     & 0.840±0.000     & 0.437±0.002     & \underline{0.247±0.002} \\
       Prediction  &  MIPE     & -     & -     & -     & \textbf{0.852±0.000}     & \underline{0.504±0.000}     & 0.225±0.002 \\
        ~  &  MIPE (AlphaFold2)     & -     & -     & -     & {0.842±0.001}     & {0.450±0.002}     & 0.213±0.003 \\
        \hline
        Joint     & \cite{ICML2020}     & 0.854±0.002     & 0.670±0.002     & 0.489±0.004      & 0.593±0.006     & 0.240±0.04     & 0.077±0.004\\
        Paratope-Epitope     & Epi-EPMP     & \underline{0.878±0.000}     & \underline{0.712±0.003}     & \underline{0.493±0.002}     & \underline{0.720±0.002}     & \underline{0.254±0.002}     & \underline{0.114±0.006} \\
        Prediction     & MIPE     & \textbf{0.894±0.000}     & \textbf{0.738±0.000}     & \textbf{0.531±0.003}     & \textbf{0.815±0.000}     & \textbf{0.422±0.001}     & \textbf{0.235±0.006} \\
        ~     & MIPE (AlphaFold2)     & {0.863±0.000}     & {0.694±0.000}     & {0.508±0.003}     & {0.795±0.000}     & {0.320±0.001}     & {0.204±0.005} \\
        \hline
    \end{tabular*}}           
    \caption{Comparison results of MIPE and baselines on three tasks. Note that the highest score in each column is in bold and the second-best score is underlined. '-' represents that the method is unsuitable for the current task.}
    \label{tab:plain}
\end{table*}

Table 1 shows the performances of MIPE and baselines in three tasks of the paratope and epitope prediction. To implement different tasks, we simultaneously train these components and two distinct predictors for joint paratope-epitope prediction, while separately training these components and a predictor for single paratope/epitope prediction. The results indicate the following observation: 
(1) For the single paratope prediction, our MIPE outperforms these baselines by an average of up to $7.1\%$ in terms of AUC and $2.0\%$-$12.9\%$ in AUPR and produces slightly lower MCC than PECAN.
(2) For the single epitope prediction, we observed a decline in overall performance compared to paratope prediction. Indeed, compared to antibodies, sequences and structures of antigens are more variable, and make the epitope prediction much more difficult \cite{akbar2021compact}. Nevertheless, our method still produces satisfying results. Compared to the sequence-based approaches BepiPred-3.0 and NetBCE, MIPE surpasses them by $0.7\%$ and $7.8\%$ in AUC, respectively. Compared with the structure-based methods PECAN and PesTo, MIPE improves AUC scores by $1.2\%$ and $21.5\%$. 
(3) For the joint paratope-epitope prediction, our method produces the best performance in terms of all metrics, particularly excelling in predicting epitopes. Compared to \cite{ICML2020} and Epi-EPMP, our MIPE exceeded their performance in the epitope prediction by 9.5$\%$ and 22.2$\%$ in AUC, 16.8$\%$ and 18.2$\%$ in AUPR, and $12.1\%$ and 15.8$\%$ in MMC. But joint paratope-epitope prediction underperforms compared to single paratope/epitope prediction, which is because the joint prediction needs the simultaneous training loss of two predictors designed for paratopes and epitopes, while single paratope/epitope prediction only one loss needs to be considered. Simultaneously constraining two distinct types of losses poses a challenge for the model, leading to performance degradation.
(4) It's worth noting that compared to the general protein binding site prediction method, PesTo, MIPE achieved better performance in both single paratope and single epitope predictions.
In summary, MIPE demonstrates superiority over the baseline methods for all three tasks, showing great promise for paratope and epitope prediction.

In practical use, the structures of antibodies and antigens are not always available, primarily due to the costs of experimental acquisition, which hinder the model from predicting binding residues for the new antibodies and antigens. Fortunately, recent years have witnessed the success of AlphaFold2, which can generate structures from sequences. Herein, we use the AlphaFold2-generated structures in the prediction process. As illustrated in Table 1, MIPE (AlphaFold2) with the generated structures still produces competitive performances in terms of various metrics compared to the baselines for three tasks and performs slightly lower than the vanilla MIPE, indicating that our model can be applicable to new antigens and antibodies without structures.

\subsection{Ablation Study}
\begin{figure}[ht]
    \center{\includegraphics[width=1.0\columnwidth]{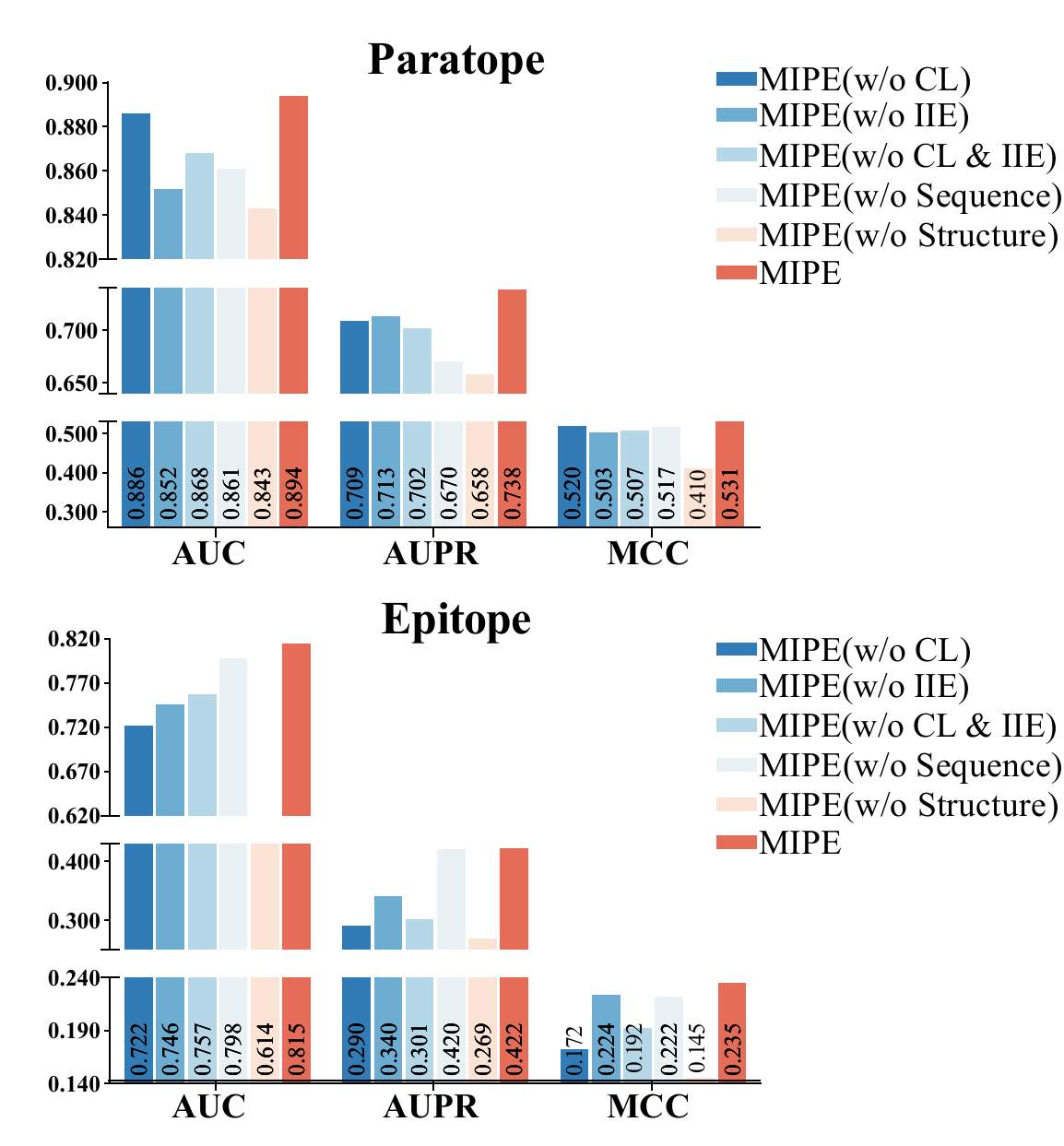}}
    \caption{Results of MIPE and its variants in joint paratope-epitope prediction.}
\end{figure}
To illustrate the necessity of each module, we conducted several comparisons between MIPE and its  variants: 
\begin{itemize}
   \item \textbf{MIPE (w/o CL)} that removes the multi-modal contrastive learning module.
   \item \textbf{MIPE (w/o IIE)} that removes the interaction informativeness estimation module.
   \item \textbf{MIPE (w/o CL \& IIE)} that removes both the multi-modal contrastive learning and interaction informativeness estimation modules.
   \item \textbf{MIPE (w/o Sequence)} that removes the sequence input.
   \item \textbf{MIPE (w/o Structure)} that removes the structure input.
\end{itemize}

Figure 3 indicates the comparisons of MIPE and its variants in joint paratope-epitope prediction (additional ablation processes can be found in Appendix 3). When removing the multi-modal contrastive learning (w/o CL), we observed a decline in all metrics. Similarly, the exclusion of both the sequence and structure encoder (w/o Sequence and w/o Structure) also resulted in a decrease across all metrics. These findings underscore the significance of leveraging multi-modal data and aligning them through contrastive learning. Additionally, the results from the variant MIPE (w/o IIE) indicate that utilizing the interaction informativeness estimation module to differentiate between binding and non-binding residues significantly contributes to the predictive performance, even in the absence of complex information. As anticipated, the performance of MIPE experienced a substantial drop when both the multi-modal contrastive learning and informativeness estimation modules were removed (w/o CL \& IIE). Overall, the above results emphasize the critical role played by these components in enhancing the model's effectiveness.

\begin{figure}[h]
\center{\includegraphics[width=1.0\columnwidth]{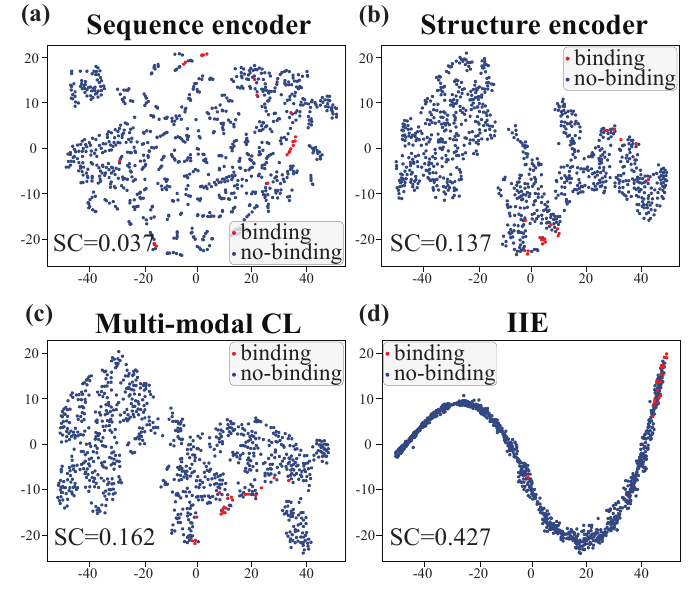}} 
\caption{The t-SNE visualization for the embeddings after sequence encoder (a), structure encoder (b), multi-modal CL (c), and interaction informativeness estimation (d).} 
\end{figure} 
To assess the efficacy of our proposed multi-modal CL and IIE, we performed t-SNE visualizations. Previous experiments have shown that our model exhibits a favorable distinction between binding and non-binding residues on antibodies/antigens, which is also reflected in the ability to discriminate these sites on each antibody/antigen effectively. 
Here, we randomly select an antigen and utilize t-SNE to visualize its embedding after the sequence encoder, structure encoder, multi-modal CL, and IIE. We utilize the Silhouette Coefficient (SC) to assess the clustering effectiveness of different embeddings. As shown in Figure 4, initially scattered embeddings after sequence and structure encoders gradually differentiate between binding and non-binding residues after multi-modal and IIE. Meanwhile, the enhanced SCs also affirm the modules' capability to extract reliable representations and discern binding and non-binding residues.

\subsection{Antibody-Antigen Binding Analysis}
\begin{figure}[h]
    \center{\includegraphics[width=1.0\columnwidth]{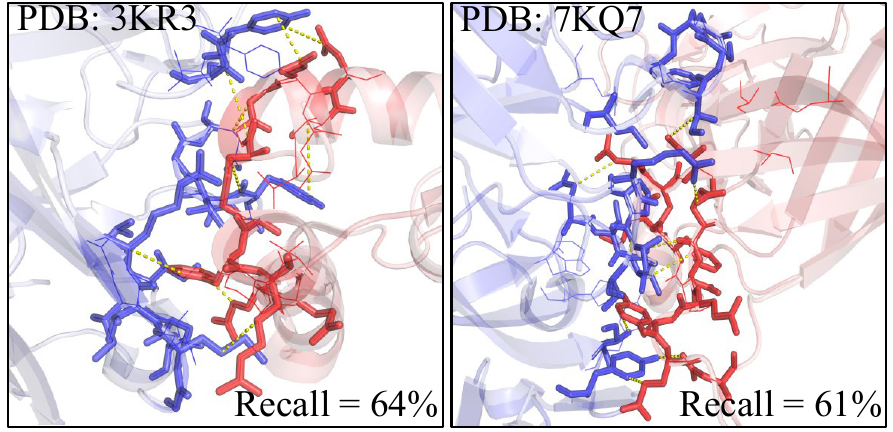}} 
    \caption{The binding visualization for antibodies and antigens, with residues from antibodies highlighted in blue and residues from antigens in red. Reference binding residues are represented by lines, while predictive binding residues are indicated by sticks. Examples of the predictive interactions are plotted in dotted lines with yellow color.
    }
\end{figure}

To further test our model's capability of capturing effective antibody-antigen interaction patterns, we randomly selected two antigen-antibody complexes (PDB: 3KR3, PDB: 7KQ7) from the PDB database as references, and employed MIPE to predict the corresponding binding residues. We visualized the actual and predicted antibody-antigen interaction maps using PyMOL. As shown in Figure 5, the left panel represents the results of the Fab antibody and insulin-like growth factor-II receptor \cite{dransfield2010human}, with 39 experimental validated binding pairs of paratopes and epitopes, and MIPE could correctly predict 24 pairs, achieving a recall score of $64\%$. Additionally, the right panel denotes the results of the Fab antibody and the Interleukin-21 receptor \cite{campbell2021combining}.
MIPE predicts 23 pairs out of 38 true pairs, yielding a recall score of $61\%$. The results show that MIPE can effectively capture the interaction patterns between antibodies and antigens, providing reliable information for the prediction.

\section{Conclusion}
In this paper, we propose a multi-modal contrastive learning and interaction informativeness estimation-based method MIPE, which uses both sequence and structure data for paratope and epitope prediction. Specifically, the multi-modal contrastive learning strategy is designed to maximize representations of binding and non-binding residues within each modality and meanwhile align uni-modal representations towards effective modal representations. Moreover, the interaction informativeness estimation module incorporates the interaction maps from the known complexes into the model and learns the specific patterns of interactions to enhance the performances. Experimental results demonstrate the effectiveness of MIPE and the importance of the above-mentioned components.

\section*{Acknowledgments}
This work was supported by the National Natural Science Foundation of China (62372204, 62072206, 61772381, 62102158); Huazhong Agricultural University Scientific $\&$ Technological Self-innovation Foundation; Fundamental Research Funds for the Central Universities (2662021JC008, 2662022JC004). The funders have no role in study design, data collection, data analysis, data interpretation, or writing of the manuscript.

\section*{Contribution Statement}
Wen Zhang, the corresponding author of this paper, supervised this project and advised on all parts of this paper. Zhiwei Wang and Yongkang Wang took the lead in designing and writing the manuscript, which should be considered to equally contribute to this work. Zhiwei Wang carried out the computational experiments, and Yongkang Wang provided help in the analysis and discussion of results. All authors reviewed and approved the final version of the manuscript. Note: in the author's list, † indicates these authors contributed equally, and * indicates the corresponding author.

\bibliographystyle{named}
\bibliography{ijcai24}

\end{document}